\documentclass[twocolumn,showkeys,amsmath,amssymb]{revtex4}

\usepackage{graphicx}
\usepackage{dcolumn}
\usepackage{bm}

\begin{document}

\title{Non-demolishing measurement of a spin qubit state via Fano resonance}

\author{V. Vyurkov$^{{\rm 1}}$}
\author{L. Gorelik$^{{\rm 2}}$}
\author{A. Orlikovsky$^{{\rm 1}}$}

\affiliation{$^{{\rm 1}}$ Institute of Physics and Technology of
the Russian Academy of Sciences, Nakhimovsky Avenue 34, Moscow
117218, Russia \\ $^{{\rm 2}}$ Chalmers University of Technology
and G\"{o}teborg University,SE-412 96 G\"{o}teborg, Sweden}

\date{\today}

\begin{abstract}
Fano resonances are proposed to perform a measurement of a spin
state (whether it is up or down) of a single electron in a quantum
dot via a spin-polarized current in an adjacent quantum wire.
Rashba-like spin-orbit interaction in a quantum dot prohibits
spin-flip events (Kondo-like phenomenon). That ensures the
measurement to be non-demolishing.
\end{abstract}


\maketitle

\section{\label{sec:level1}Introduction}
The solid state structures seem quite promising to implement
quantum computers. The first proposal of the solid state quantum
computer based on quantum dots was put forward in 1998 by D. Loss
and P. DiVincenzo \cite{loss}. The quantum calculations were
offered to be performed on single electron spins placed in quantum
dots. To read out the result one should measure the state of the
quantum dot register consisting of single spins. Several
possibilities were proposed there. In particular, two of them
exploited a spin blockade regime. One could use spin dependent
tunneling of a target electron into a quantum dot with a definite
spin orientation of a reference electron. The final charge state
of the dot (whether tunneling occurred or not) could be tested by
a single electron transistor (SET) quite sensitive for an
environment charge.

In the same year Kane offered a solid state quantum computer based
on $^{{\rm 3}{\rm 1}}$P atoms embedded in Si substrate
\cite{kane}. The computation was to be performed on $^{{\rm 3}{\rm
1}}$P nuclei spins mediated by outer shell electrons. The
inventive procedure to transfer a resulting nucleus spin state
into an electron spin state was proposed. The latter could be
measured by single electron tunneling into the reference atom,
i.e. almost with the same means as that in Ref. \cite{loss}. This
idea was much developed in Ref. \cite{recher} where it was
proposed to test a single electron spin in a quantum dot by a
spin-polarized current passing through this dot sandwiched between
leads. The inspiring experiment demonstrating such a readout of a
single electron spin placed in an open quantum dot by detecting a
current passing through the dot was presented in \cite{giorga}.

Far earlier than the first solid state quantum computer implementations 
were put forward a set of scanning tunneling microscopy experiments 
to observe an evolution of a single spin with the help of spin-polarized 
current emanating from a magnetic tip had begun \cite{S1, S2, S3, S4}.

Most of proposals of spin-based quantum computers relate to
electron spins although their relaxation is much faster than that
of nucleus spins. It is caused by relatively easier and faster
operation upon them and a possibility of measurement of individual
electron spin state.

   In general, most of proposals of spin-based quantum computers 
relate just to electron spins although their relaxation is much faster 
than that of nucleus spins. It is caused by relatively easier and 
faster operation upon them and a possibility of measurement of 
individual electron spin state.    All suggestions of individual 
electron spin measurement made so far are based on exchange 
interaction between a target electron and a referenced electron 
which spin orientation is known. 

One of the publications on the topic \cite{engel} concerns the
problem how to perform a particular measurement of spin states via
tunneling in adjacent quantum dots, namely, to make clear whether
spins are parallel or antiparallel. This kind of measurement could
be employed for quantum calculations instead of organizing an
interaction between qubits which can hardly be controlled with
fairly high accuracy. The recent paper \cite{sarovar} discusses a
possibility to test a single spin via spin-dependent scattering
inside a field effect transistor channel. However, spin-flip
(Kondo-like) phenomena were unreasonably ignored there.

Here we examine an opportunity to employ a quantum wire with 
a spin-polarized current to measure a single electron spin inserted 
in a nearby quantum dot. Firstly, in Ref. \cite{vyurkov1}
 there was pointed out 
to an opportunity that a current passing through a quantum wire 
could be quite sensitive to a nearby charge due to Coulomb blockade. 
Further the sensitivity of a quantum wire with spin polarized current 
to a spin state of an electron in adjacent quantum dot was clarified 
in Ref. \cite{vyurkov2}. 
There the possibility of spin blockade of current was 
elucidated. In this paper we propose to employ Fano resonances 
in a quantum wire to make the measurement much more sensitive. 
We also regard the question how to make a non-demolishing 
measurement by virtue of a spin-orbit interaction.  

\section{\label{sec:level1}Quantum wire to measure a single spin state}
We suggest a measurement which allows via detecting the current to
conclude whether the electron in a quantum dot has the same spin
orientation as electrons in the quantum wire or opposite one.

\begin{figure*}
\includegraphics{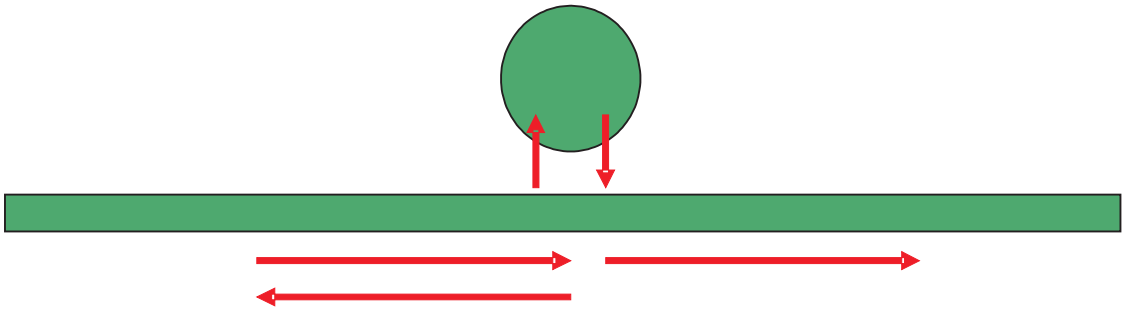}
\caption{\label{fig:wide} A quantum wire coupled to a nearby
quantum dot via tunneling.}
\end{figure*}

The structure under consideration is schematically depicted in
Fig. 1. It consists of a quantum wire transmitting a spin
polarized current coupled to a nearby quantum dot via tunneling.
The current through the wire is determined by well-known Landauer
formula

\begin{equation}
\label{eq1} I(V_{sd} ) = {\frac{{e}}{{h}}}{\sum\limits_{i =
0}^{\infty}  {}} \int {dE \ T_{i} (E){\left[ {f_{s} (E) - f_{d}
(E)} \right]}} ,
\end{equation}

\noindent where summation is fulfilled over all modes of
transversal quantization in the wire, $T_{i}(E)$ is a transmission
coefficient for i-th mode dependent on the total energy $E$, the
factor $e/h$ arises from a conductance quantum $e^{2}/h$ for spin
polarized current in a ballistic wire, $f(E)$ is Fermi-Dirac
distribution function

\begin{equation}
\label{eq2} f(E) = {\frac{{1}}{{1 + \exp {\frac{{E - \mu}}
{{kT}}}}}},
\end{equation}

\noindent providing the chemical potentials in the source contact
$\mu _{{\rm s}}$ and in the drain contact $\mu _{{\rm d}}$ are
shifted by bias: $\mu _{{\rm s}}-\mu _{{\rm d}}=e V_{{\rm s}{\rm
d}}$. The transmission coefficients $T_{i}(E)$ may be different
for different spin state of the quantum dot electron with respect
to spin polarization of electrons in quantum wire. This results in
different current. The measurement is operated by potentials on
gate electrodes.

Recently the proposals appeared to exploit Fano resonances for a
measurement of an individual electron spin state with the help of
quantum wire or quantum constriction which in fact could be
regarded as a merely short quantum wire \cite{mourokh,vyurkov3}.
Indeed, Fano resonances make such a measurement more sensitive.

Fano effect in the considered structure means the following. An
electron moving along the quantum wire can partially penetrate
into the quantum dot due to tunneling. The interference between
two routes, one of which passes through the dot and another
doesn't, determine the transmission coefficient of an electron
through the wire. In other words, the discrete energy spectrum in
the dot interferes with a continuum in the wire. This interference
becomes destructive when the energy of an electron in wire
coincides with that in a dot. This results in backscattering which
can be detected by a dip on a current-voltage curve, so called
Fano antiresonance.

Transmission coefficient $T$ for an electron with the energy
detuning from the resonance \textit{$\varepsilon $} is supplied by
the expression

\begin{equation}
\label{eq0} T = {\frac{{{\left| {\varepsilon + q\Gamma}
\right|}^{2}}}{{\varepsilon ^{2} + \Gamma ^{2}}}},
\end{equation}

\noindent where $\Gamma$ stands for the level broadening and $q$
is the Fano asymmetry factor. In general $q$ is a complex number
depending on scattering and relaxation in the system. For a
ballistic quantum wire and a quantum dot without relaxation $q
\approx 0$. Here we assume that this is the case.

However, the model put forward in \cite{mourokh} does not take
into account the possibility of spin flip, i.e. Kondo-like
phenomenon. Indeed, the spin exchange between an incident electron
and an electron in quantum dot takes the same time required for an
incident electron to be backscattered. Therefore, the initial spin
state of the measured electron becomes demolished after only one
incident electron is backscattered. Here we endeavor to circumvent
such a shortcoming introducing a spin-orbit interaction into the
system.

\begin{figure*}
\includegraphics{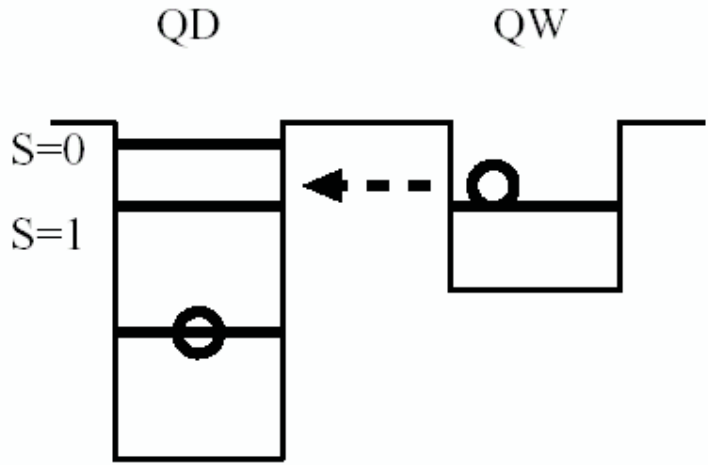}
\caption{\label{fig:wide} An electron tunnels from a quantum wire
(QW) to the excited state in a quantum dot (QD) with total spin
$S=1$.}
\end{figure*}

However, for the sake of clarity we firstly discard a spin-orbit
interaction to highlight the disadvantages of such an approach to
measurement. The general way of measurement looks like follows. A
conducting electron (an electron at the Fermi level) in the
quantum wire tunnels to an excited state in a quantum dot which is
split with respect to the total spin $S$ because just a total spin
determines the exchange energy. We also suppose that the state
$S=1$ has a smaller exchange energy as the state $S=0$ (Fig. 2).
This is valid, at least, for the excited state with orbital moment
$L=1$. It should be noted that the Lieb-Mattis theorem is not
valid for excited states. This theorem only claims that the ground
state definitely has $S=0$ for even number of electrons. Anyhow,
this suggestion is not crucial for the measurement.

We suppose that the initial state of the system is the following.
For the sake of definiteness, the spin of electron in quantum wire
is always directed up ($\uparrow $) and the spin of target
electron in the ground state of quantum dot is directed up
($\uparrow$) or down ($\downarrow$). Evidently, only mutual spin
orientation of both electrons matters. Due to tunneling this state
evolves into some excited state of two electrons in the quantum
dot.

The Hamiltonian describing tunneling of an electron from the
quantum wire into an exited level in the quantum dot in absence of
spin-orbit interaction reads

\begin{equation}
\label{eq3} H_{0} = \varepsilon _{0} + (\varepsilon _{1} + U_{C} -
J\vec {S}_{0} \vec {S}_{1} )a_{1}^{ +}  a_{1} +  
\\{\sum\limits_{j}
{}} \{T_{j} (a_{1}^{ +}  b_{j} + b_{j}^{ +}  a_{1} ) + \varepsilon
_{j} b_{j}^{ +}  b_{j}\},
\end{equation}

\noindent where $\varepsilon _{{\rm 0}}$ is a ground level energy
of a single electron in the quantum dot, $\varepsilon _{{\rm 1}}$
is an excited level energy, $U_{C}$ is a direct Coulomb
interaction between an electron in the ground state and an
electron in the excited state, $a_{1}^{ +}$ and $a_{1}^{}$ are
operators of creation and annihilation of an electron in the
excited level in the quantum dot, $b_{j}^{ +}$ and $b_{j}$ are
operators of creation and annihilation of an electron in the
quantum wire in j-th state, in general, the longitudinal momentum
and the transversal quantization subband (mode) number are
ascribed to the index $j$, $J$ is the strength of exchange
interaction, $\vec {S}_{0}$ and $\vec {S}_{1}$ are spin operators
for an electron in the ground level and that in excited level,
respectively. In general, the sign of exchange energy (therefore,
the sign of $J)$ could be positive or negative. It is only known
for sure that the ground state of the system composed of two
electrons definitely corresponds to the total spin $S=0$ owing to
Lieb-Mattis theorem.

It is convenient to present the spin Hamiltonian in the form

\begin{equation}
\label{eq4} \vec {S}_{0} \vec {S}_{1} = S_{0}^{z} S_{1}^{z} +
S_{0}^{ +}  S_{1}^{ -} ,
\end{equation}

\noindent where $S_{0}^{z}$ and $S_{1}^{z}$ are operators of
$z$-projection, $S_{0}^{ +}$ and $S_{1}^{ -}$ are raising and
lowering operators of $z$-projection. The second term describes
spin-flip processes, i.e. Kondo-like phenomena.

Hereafter, we incorporate into consideration only three lowest
states of two electrons in the quantum dot: the ground state with
total spin $S=0$ and two excited states corresponding to the same
space wave function but the different total spin $S=0$ and $S=1$.
The energies of these excited states differ due to exchange
interaction.
 
We suppose that the tunneling coupling $T$ of both
states with a quantum wire continuum is the same as the space wave
functions are the same. Hereafter, we employ the designations
analogous to that in Ref. \cite{engel}.

There are two triplet states with $S=1$ which can originate after
an electron with spin up tunnels from a quantum wire into a
quantum dot

\begin{equation}
\label{eq-1} \uparrow _{D} \uparrow _{D}
\end{equation}

\noindent for parallel spins and

\begin{equation}
\label{eq-2}{{{\left[ { \uparrow _{D} \downarrow _{D} + \downarrow
_{D} \uparrow _{D}} \right]}} \mathord{\left/ {\vphantom {{{\left[
{ \uparrow _{D} \downarrow _{D} + \downarrow _{D} \uparrow _{D}}
\right]}} {\sqrt {2}}} } \right. \kern-\nulldelimiterspace} {\sqrt
{2}}}
\end{equation}

\noindent for antiparallel spins. There could also occur a singlet
state with $S=0$

\begin{equation}
\label{eq-3}{{{\left[ { \uparrow _{D} \downarrow _{D} - \downarrow
_{D} \uparrow _{D}} \right]}} \mathord{\left/ {\vphantom {{{\left[
{ \uparrow _{D} \downarrow _{D} - \downarrow _{D} \uparrow _{D}}
\right]}} {\sqrt {2}}} } \right. \kern-\nulldelimiterspace} {\sqrt
{2}}}
\end{equation}

\noindent for antiparallel spins. The above formulae are sensitive
to position: the first place corresponds to the ground state.

Tunneling to the ground state from a quantum wire seems
inappropriate for the measurements because the target electron may
escape from the quantum dot the same way as the reference
electron.

Therefore, tunneling to the excited state is preferable as it
leaves the target electron in the dot.

First of all we must discuss the initial state of the system when
only a target electron is situated in the ground state in a
quantum dot and there is a reference electron in a quantum wire.

For parallel spin orientation

\begin{equation}
\label{eq-4} \uparrow _{D} \uparrow _{W}
\end{equation}

This state relates to a definite value of total spin $S=1$ and
there is no entanglement between spin and space coordinates.

For antiparallel spins the state is

\begin{equation}
\label{eq-5} \downarrow _{D} \uparrow _{W} = {{{\left[ {
\downarrow _{D} \uparrow _{W} - \uparrow _{D} \downarrow _{W}}
\right]}} \mathord{\left/ {\vphantom {{{\left[ { \downarrow _{D}
\uparrow _{W} - \uparrow _{D} \downarrow _{W}} \right]}} {\sqrt
{2}}} } \right. \kern-\nulldelimiterspace} {\sqrt {2}}}  +
{{{\left[ { \uparrow _{D} \downarrow _{W} + \downarrow _{D}
\uparrow _{W}} \right]}} \mathord{\left/ {\vphantom {{{\left[ {
\uparrow _{D} \downarrow _{W} + \downarrow _{D} \uparrow _{W}}
\right]}} {\sqrt {2}}} } \right. \kern-\nulldelimiterspace} {\sqrt
{2}}}
\end{equation}

Indeed, this state is a superposition of $S=0$ and $S=1$,
moreover, there exists an entanglement of space and spin
coordinates. Worth noting both space components of this state do
not interfere because they are orthogonal with respect to total
spin.

When the conditions shown in Fig. 2 are effectuated and electrons
from the Fermi level in quantum wire can tunnel to the state with
$S=1$ in a quantum dot the reflection coefficient for the
antiparallel spins (10) is exactly twice smaller than that for
parallel spins (9). This looks like the basis of a spin state
measurement.

Actually, it is an illusion that foregoing procedure is already
satisfactory for the measurement. Really, Kondo-like phenomena,
i.e. spin-flip events should be involved into consideration. The
spin-flip occurs within the period of time $\tau \sim (J / \hbar
)^{ - 1}$. The broadening of levels with $S=0$ and $S=1$ depends
on tunneling rate, so that $\Gamma \approx T$. The claim for these
levels to be clearly distinguished requires $J \gg T$. It follows
that the spin flip process must be much faster than tunneling.

We propose to employ spin-orbit interaction inside a quantum dot
to prevent spin-flip. Hereafter, we analyze the simplest case when
a coin-like quantum dot is formed of a quantum well. It is
sketched in Fig. 3. The arrows there indicate the interface
electric field at different interfaces caused by conduction (or
valence) band discontinuity. For instance, this kind of dot could
be fabricated by etching with a subsequent overgrowing of a host
semiconductor with a wider gap. Rashba spin-orbit interaction
\cite{rashba} originates in interface electric field and strong
conduction and valence bands coupling in narrow-gap semiconductors
\cite{zakharova,wissinger}. This mechanism of spin-orbit
interaction turned out to be several orders stronger than that
based on electric field in one-band model. It seems as the most
suitable to get the goal of non-demolishing measurement of a spin
qubit state.

The original Rashba Hamiltonian describing spin-orbit interaction
in a two-dimensional electron gas (2DEG) reads \cite{rashba}:

\begin{equation}
\label{eq5} \hat {H}_{R} = \alpha _{R} [\hat {\vec {S}}\times \hat
{\vec {k}}]\vec {\upsilon}  \equiv \alpha _{R} [\hat {\vec
{k}}\times \vec {\upsilon} ]\hat {\vec {S}},
\end{equation}

\noindent where $\alpha _{{\rm R}}$ is Rashba constant, $\vec {k}
= - i{\frac{{\partial}} {{\partial \vec {r}}}}$ is an operator of
in-plane moment, $\vec {S}$ is a spin operator, and a unit vector
$\vec {\upsilon}$ is directed perpendicular to 2DEG plane. The
Rashba constant $\alpha _{{\rm R}}$ is non-zero for a 2DEG in
non-symmetric quantum well.\textbf{ }Unfortunately, Rashba
Hamiltonian (\ref{eq5}) results in an entanglement of spin and
space variables for an electron state in a quantum dot cut of a
quantum well. It occurs even in a ground state. This kind of a
quantum dot is not suitable for a spin qubit application. The
appropriate quantum dot should be made of a symmetric quantum well
with zero original Rashba term (\ref{eq5}).

Rashba Hamiltonian (\ref{eq5}) is widely used for a 2DEG
originating at a unique interface in a heterostructure, for
example, at common GaAs/AlGaAs interface. We adopt this
description to introduce a spin-orbit interaction caused by
interface field at side wall of the dot. To that end, we propose a
model Rashba-like Hamiltonian

\begin{equation}
\label{eq6} \hat {H}_{RS} = \beta [\hat {\vec {S}}\times \hat
{\vec {k}}]\hat {\vec {r}} = \beta [\hat {\vec {k}}\times \hat
{\vec {r}}]\hat {\vec {S}} = \beta \hat {\vec {L}}\hat {\vec {S}},
\end{equation}

\noindent where $\hat {\vec {L}} = [\hat {\vec {k}}\times \hat
{\vec {r}}]$ is an operator of angular momentum of an electron in
the dot which appears after permutation, $\hat {\vec {S}}$ is a
spin operator, $\beta $ is a coefficient analogous to Rashba
constant, it could be roughly evaluated as $\beta \approx \alpha
_{{\rm R}}/D$, where D is a dot diameter.

\begin{figure}
\includegraphics{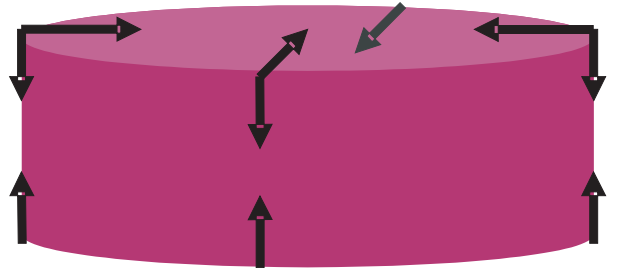}
\caption{\label{fig:epsart} A coin-like quantum dot formed of a
symmetric quantum well. Arrows indicate the interface electric
field.}
\end{figure}

\begin{figure*}
\includegraphics{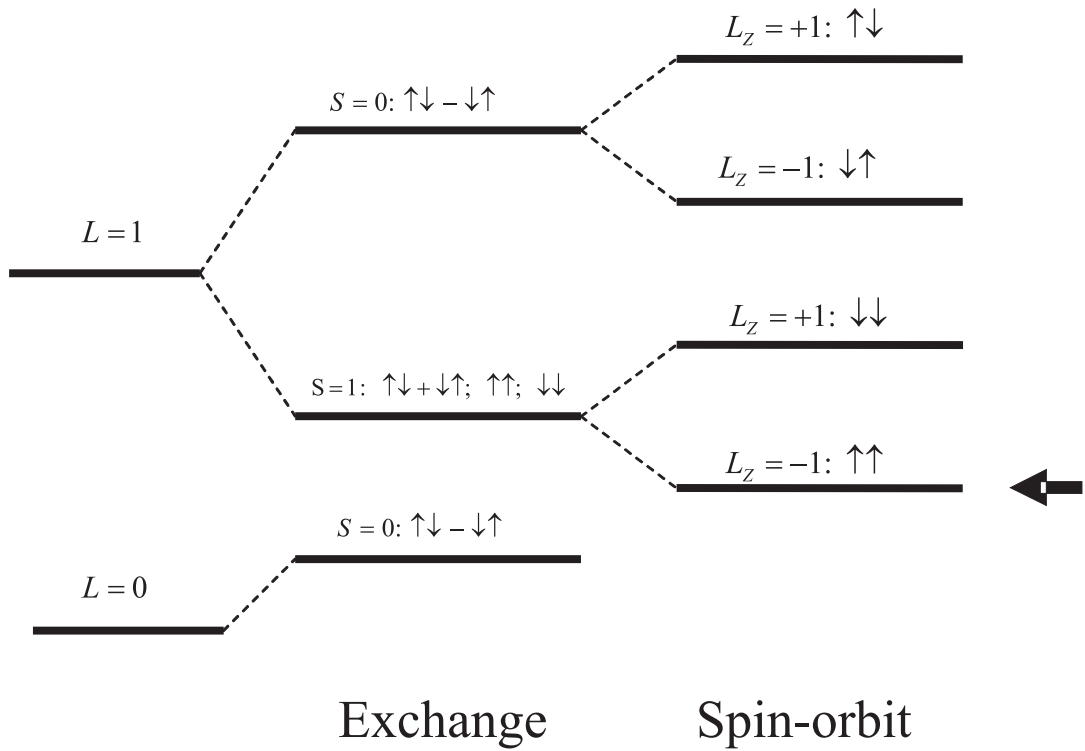}
\caption{\label{fig:wide} Two-electron energy levels in a quantum
dot split by exchange and spin-orbit interaction. The arrow
indicates the level to which an electron tunnels from a quantum
wire.}
\end{figure*}

For the measured electron in a ground state with L=0 spin-orbit
interaction equals zero. The electron can possess any orientation
of spin. Therefore, this a good quantum dot for a spin qubit
manipulation. For the excited state with angular momentum $L=1$
the Hamiltonian (\ref{eq6}) looks like

\begin{equation}
\label{eq7} H_{RS} = \beta L_{1Z} S_{1Z} ,
\end{equation}

\noindent where $L_{1Z} = \pm 1$ is a z-projection of orbital
moment, $S_{1Z} = \pm 1 / 2$ is z-projection of electron spin in
the excited state. The spin-orbit term (\ref{eq5}) should be added
to the Hamiltonian (\ref{eq3}) to acquire the resultant
Hamiltonian

\begin{equation}
\label{eq8} H = H_{0} + H_{RS} .
\end{equation}

The eigen-states of the Hamiltonian (\ref{eq8}) relevant to two
electrons in the dot are depicted in Fig. 4. Tunneling of an
electron from a quantum wire to a quantum dot occurs if only the
latter has the same spin orientation as the measured electron in
dot (the arrow in Fig. 4 marks the proper level).

Spin-orbit splitting makes impossible the spin-flip due to energy
conservation law. As it was reported in Ref. \cite{nitta} Rashba
spin-orbit splitting in $A_{{\rm I}{\rm I}{\rm I}}B_{{\rm V}}$
heterostructures may attain to several meV. At the same time, the
Zeeman energy splitting proposed in Ref. \cite{engel} for the same
reason to suppress spin-flip processes in a dot approximately
equals $0.3 meV$ even in rather big magnetic field $5T$.

In accordance with expression (3) the mean reflection coefficient
$R=1-T$ in the range $-\Gamma<\varepsilon<+\Gamma$ is
approximately $1/3$. It means that the relative decrease in
current for parallel spins is around $1/3$ if the bias $V \le
\Gamma/e$. For antiparallel spins it is twice smaller, i.e. around
$1/6$. For the sake of better sensitivity the optimal bias $V$ for
a ballistic quantum wire should be chosen around $V=\Gamma/e$.
Then the absolute value of current could be roughly estimated for
a single mode quantum wire as $I=G_{0}V$, where $G_{0}=e^{2}/h=(26
kOhm)^{{\rm -} {\rm 1}}$ is a conductance quantum for
spin-polarized current in the wire. The possible level broadening
$\Gamma $ is restricted only by spin-orbit splitting. Supposing
the latter as several meV we are able to choose the broadening as
1meV. Substituting $V=1mV$ one arrives at the current equal to
$4\cdot10^{{\rm -} {\rm 8}} A$ which could be easily measured by
up to date equipment. Moreover, this current exceeds that in a
single electron transistor (SET). The greater is the current the
faster is its measurement. One more significant advantage of a
quantum wire is that it can be emptied during computing and,
therefore, unlike to a SET the quantum wire does not introduce an
additional decoherence in the system that time.

Worth noting a quantum wire allows to perform a partial
measurement of the state of two adjacent spin qubits like in
\cite{engel} or even distant ones: whether they are parallel or
antiparallel. This also provides a possibility of quantum
computation without organizing a perfectly controllable
interaction between qubits.

If for some reason the reflection coefficient is too low,
hopefully, the sensitivity may be augmented when N identical
qubits are placed in series along the wire. When qubits are
situated randomly and, therefore, interference does not matter the
sensitivity may rise as $\sim N$. When there is an order in qubit
positions and the interference is significant the sensitivity
increases as $\sim N^{{\rm 2}}$.

The opportunity to perform the proposed measurement is confirmed
by findings in Ref. \cite{sato}. Schematically the structure was
almost the same as that in Fig. 1. The combined Fano-Kondo
anti-resonances were observed in the $I-V$ curve and exploited to
test relaxation in multi-electron quantum dot. In principle, this
set up could serve as a prototype of our proposal.

\section{\label{sec:level1}Conclusion}
We have examined the possibility to use Fano-Rashba resonances for
non-demolishing measurement of a spin state (whether it is up or
down) of a single electron in a quantum dot (spin qubit) via a
spin-polarized current in an adjacent quantum wire. The spin-orbit
interaction in a quantum dot prohibits spin-flip events
(Kondo-like phenomenon). That makes the measurement
non-demolishing.

\begin{acknowledgments}
The research was supported by NIX Computer Company
(science@nix.ru), grant F793/8-05, via the grant of The Royal
Swedish Academy of Sciences, and also by Russian Basic Research
Foundation, grants \# 08-07-00486-a-a and \# 06-01-00097-a.
\end{acknowledgments}

\bibliography{non_demolishing_5}

\end{document}